\pdfoutput=1
\documentclass[10pt,twocolumn]{article} 
\usepackage{csquotes}
\usepackage[british]{babel}
\usepackage{graphicx}
\graphicspath{{figures/}}
\DeclareGraphicsExtensions{.pdf,.jpeg,.png}
\usepackage{amsmath}
\usepackage{algorithmic}
\usepackage{url}
\usepackage{wrapfig}
\usepackage[nolist]{acronym}
\usepackage{subcaption}
\usepackage{fixltx2e}

\begin{document}
\author{Peter van Stralen\\
Insititute for Informatics, University of Amsterdam\\
p.vanstralen@uva.nl}

\title{Using Chip Multithreading to Speed Up\\Scenario-Based
Design Space Exploration \thanks{We would like to thank Bart Muijzer
and Ruud van der Pas from Oracle
Netherlands for their support during the usage of the SPARC T3-4.}
}
\date{} 

\maketitle

\begin{abstract}
To cope with the complex embedded system design, early design space
exploration (DSE) is used to make design decisions early in the design
phase. For early DSE it is crucial that the running time of the
exploration is as small as possible. In this paper, we describe both the
porting of our scenario-based DSE to the SPARC T3-4 server and the
analysis of its performance behavior.
\end{abstract}

\begin{acronym}[LLLLLL] 
\acro{ASIC}{Application Specific Integrated Circuit}
\acro{ASIP}{Application Specific Instruction-set Processor}
\acro{CMP}{Chip Multiprocessor}
\acro{CMT}{Chip Multithreading}
\acro{DSE}{Design Space Exploration}
\acro{DMR}{Double Modular Redundancy}
\acro{FG-MT}{Fine-Grained Multithreading}
\acro{FCFS}{First-Come-First-Serve}
\acro{FIFO}{First-In-First-Out}
\acro{FIT}{Failures In Time}
\acro{FPGA}{Field Programmable Gate Array}
\acro{FS}{Feature Selection}
\acro{GA}{Genetic Algorithm}
\acro{HYB}{Hybrid Approach}
\acro{KPN}{Kahn Process Network}
\acro{MJPEG}{Motion-JPEG}
\acro{MOEA}{Multi-Objective Evolutionary Algorithm}
\acro{MOP}{Multi-Objective Optimization Problem}
\acro{MPI}{Message Passing Interface}
\acro{MPSoC}{Multi-Processor System-on-Chip}
\acro{MoC}{Model of Computation}
\acro{MTTF}{Mean Time To Failure}
\acro{NBTI}{Negative Bias Temperature Instability}
\acro{OWP}{Outside World Process}
\acro{RFCS}{Roll Forward Checkpointing Scheme}
\acro{RPC}{Remote Procedure Call}
\acro{RTL}{Register Transfer Level}
\acro{SAFE}{Sesame Automated Fault-tolerant Explorer}
\acro{SBS}{Sequential Backward Selection}
\acro{Sesame}{Simulation of Embedded Systems Architectures for
Multi-level Exploration}
\acro{SFS}{Sequential Forward Selection}
\acro{SoC}{System-on-Chip}
\acro{SS}{Stable Storage}
\acro{SUE}{Single Upset Event}
\acro{SWIFI}{SoftWare Initiated Fault Injection}
\acro{TDMA}{Time Division Multiple Access}
\acro{TLB}{Translation Lookaside Buffer}
\acro{TMR}{Triple Modular Redundancy}
\acro{YML}{Y-chart Modeling Layer}
\acro{QoS}{Quality-of-Service}
\end{acronym}

\section{Introduction}
A significant amount of research has been performed on system-level
\ac{DSE} for \acp{MPSoC}s \cite{gries04, pimentel06, jia10} during
the last two decades. The majority of this work is focused on the analysis of MPSoC
architectures under a single, static application workload. The current
trend, however, is that application workloads executing on embedded
systems become more and more dynamic.

\begin{figure}[t!]
\centering
\includegraphics[width=2.4in]{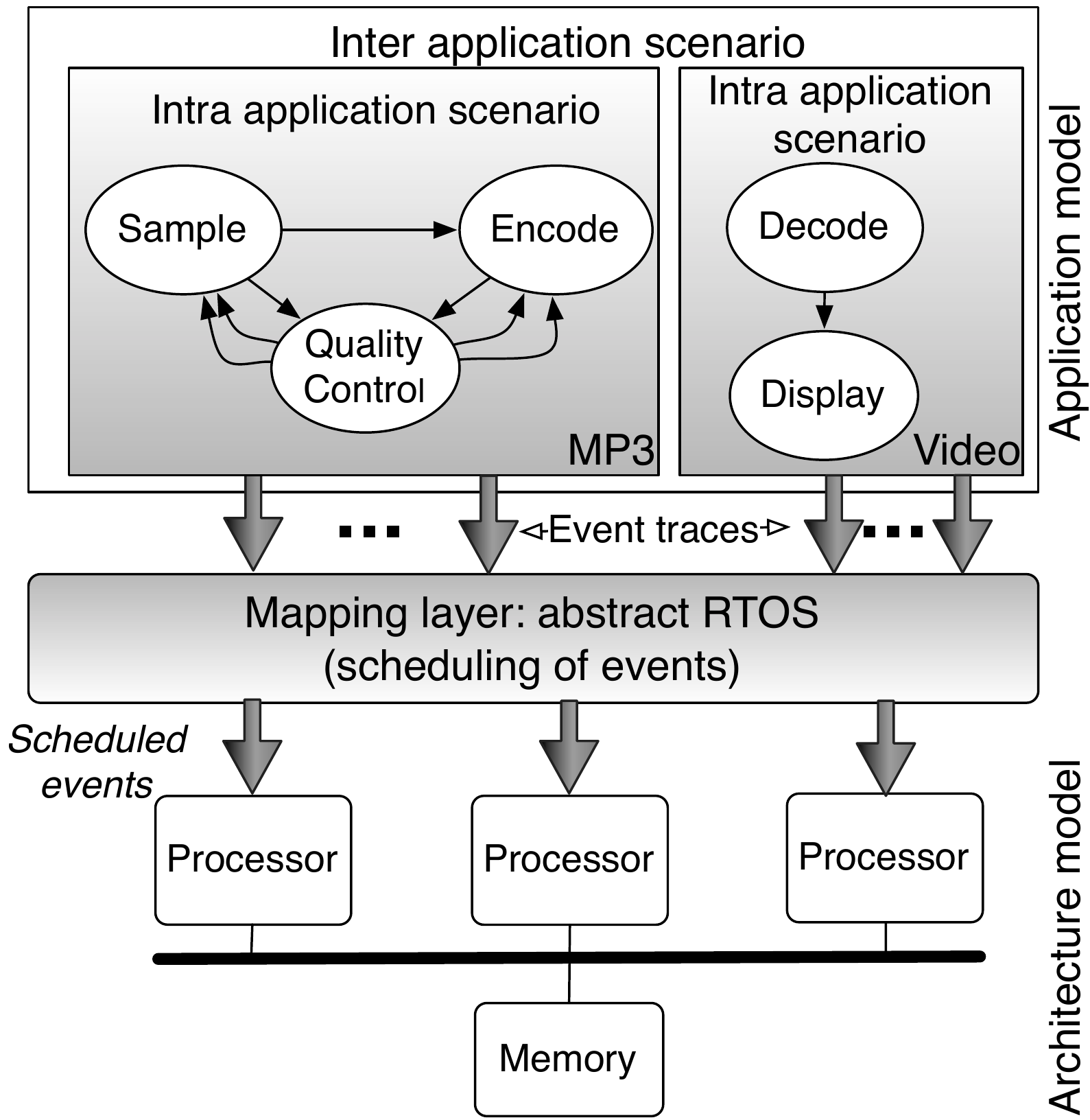}
\caption{High level scenario-based MPSoC simulation}
\label{fig:sesame}
\end{figure}

Recently, we have introduced the scenario-based \ac{DSE} environment
\cite{stralen10c} that exploits workload scenarios \cite{gheorghita09}
to model both the dynamism between and within the applications that are
mapped onto a MPSoC. As a basis for scenario-based \ac{DSE}, a
scenario-aware version of our high level simulation framework Sesame
\cite{pimentel06, stralen10b} is used.  Within scenario-aware Sesame
(as illustrated in Figure \ref{fig:sesame}) a separation of concerns
is used with separate models for the application, the architecture and
the mapping. In the application model the functional behavior of the
application is described using intra- and inter-application scenarios.
Intra-application scenarios describe the
dynamic behavior within an application, whereas the inter-application
scenarios describe the dynamic behavior between multiple applications
(i.e., which applications can run concurrently).  The structure of the
applications themselves are described using Kahn process networks
\cite{kahn74}. Next, the architecture model describes the
non-functional behavior (e.g., used power, elapsed cycles) of the
MPSoC design. To connect the application model and the architecture
model, the mapping layer maps both the processes and communication
channels in the applications onto a component in the architecture.

\begin{figure}[t!]
\centering
\includegraphics[width=3in]{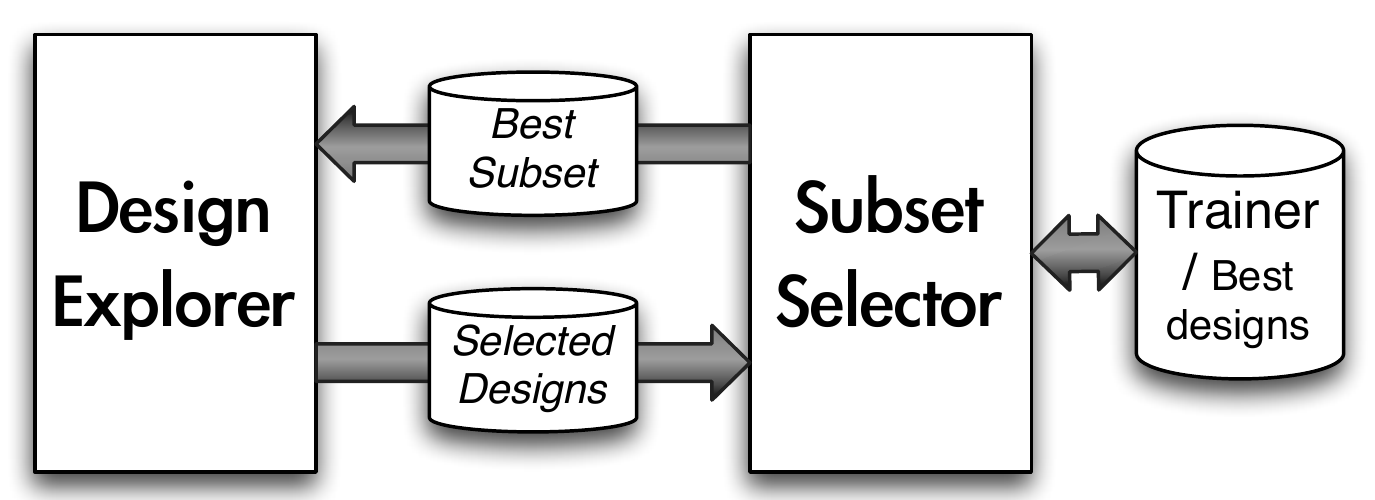}
\caption{The framework for
scenario-based DSE using feature selection.}
\label{fig:sdse}
\end{figure}

A complicating factor of embedded system design, however, is that
there is an exponential number of potential mappings. That is why a
search algorithm is required to efficiently explore the design space.
Therefore, our scenario-based \ac{DSE} framework aims at an efficient search for (sub-)optimal
mappings of embedded systems with dynamic multi-application
workloads. As shown in Figure \ref{fig:sdse}, the
framework consists of two components: the design explorer and the
subset selector. In the design explorer, there is searched for the
optimal MPSoC mapping. For this purpose, a genetic algorithm
\cite{mitchell98} is used that applies natural evolution on a population of
mappings to identify high quality mappings. This quality is determined
by simulating each of the mappings in the population. Although a
Sesame simulation typically takes less than a second, there are many
mappings that need to be evaluated. On top of that, each mapping needs
to be evaluated for multiple application scenarios. To speed up the
evaluation of a single mapping, a representative subset of application
scenarios is used. This mapping is identified by the subset selector.
Based on a set of training mappings (that is based on selected designs
from the design explorer), the subset selector dynamically selects a
representative scenario subset. Since, the representative subset of
scenarios is dependent on the current designs in the design explorer,
both the design explorer and the subset selector are running
simultaneously.

To efficiently run the scenario-based \ac{DSE}, we have ported it to the
SPARC T3-4 server and studied its performance behavior. With $512$ hardware threads, the SPARC T3-4
server is perfectly suitable to be used for the embarrassingly
parallel scenario-based \ac{DSE}. During the search of the scenario-based
\ac{DSE}, a number of worker threads are used to perform Sesame simulations
in parallel. Over time, each worker thread fires many processes to
perform simulation jobs that can be execute on the hardware threads of
the SPARC T3-4.

In the remaining sections of this paper, we start by discussing
the SPARC T3-4 server in more detail. Next, Section \ref{sec:impl}
describes the unoptimized implementation of the scenario-based \ac{DSE}.
The following two sections (Section \ref{sec:profSesame} and
\ref{sec:profSDSE}) will shows how Sesame and the scenario-based \ac{DSE}
are profiled to optimize the final design of the scenario-based \ac{DSE}
(Section \ref{sec:final}). After that, experiments show both the
performance and the bottlenecks of the scenario-based framework.
Finally, a conclusion is given.

\section{The SPARC T3-4}\label{sec:sparc}
\begin{figure}[th]
\centering
\includegraphics[width=3.25in]{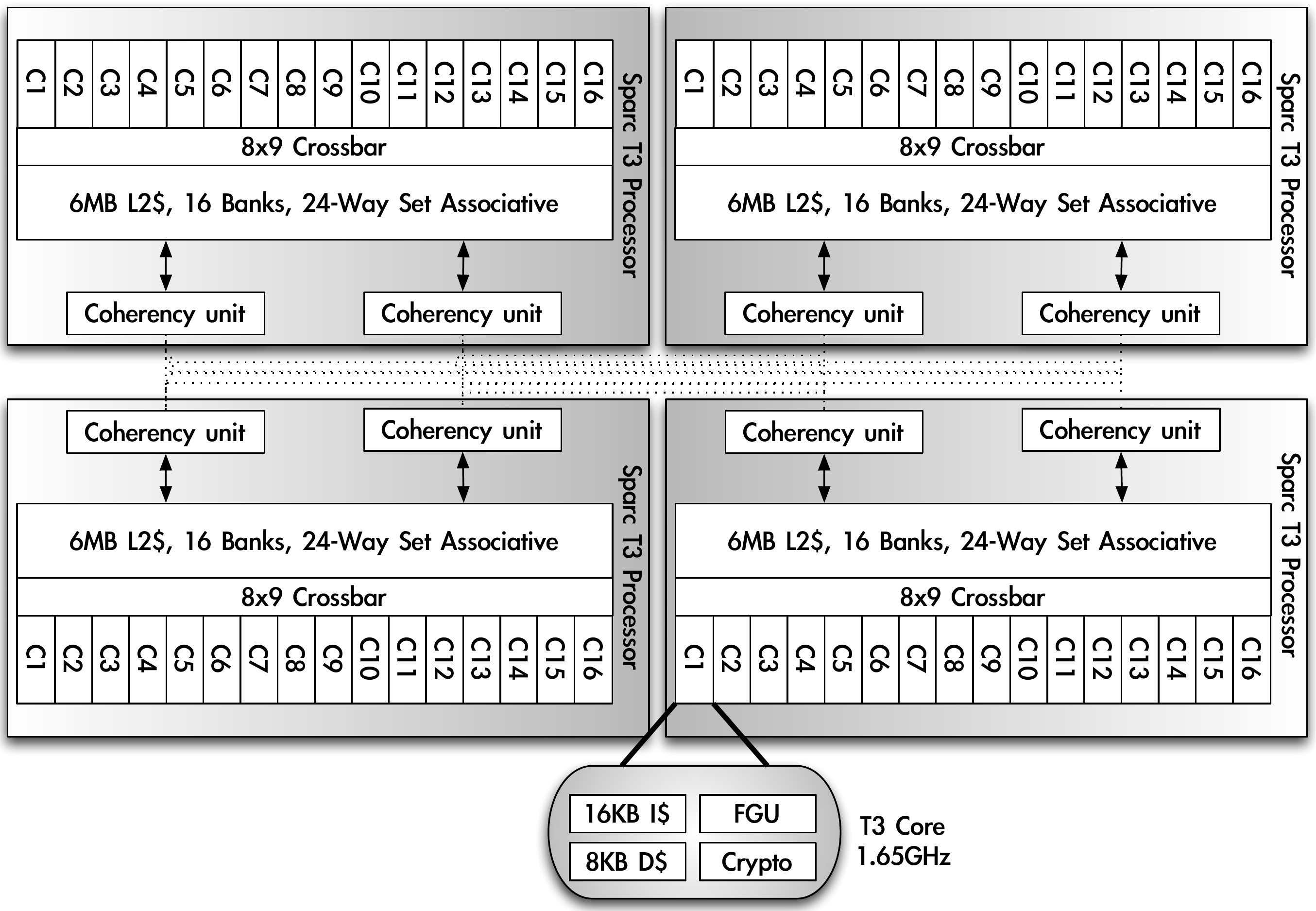}
\caption{A schematic view of the SPARC T3-4 processor.}
\label{fig:t3:overview}
\end{figure}

Most of the multi-core processors are implemented as a \ac{CMP}. On a
single chip, multiple (identical) cores are placed to enable thread
level parallelism. Although thread level parallelism improves the
number of potential operations per second, it does not deal with the
memory gap. The memory gap is the disparity between the processor
speed and the memory speed and, due to this disparity, the memory
latency is an important aspect of the processor performance. If the
processor is not able to find alternative work while an application is
blocked on a memory call, both the performance and energy consumption
of a processor degrades.

The SPARC T3 processor \cite{oracleT3} tries to minimize the idle time of the
processor by using \ac{CMT}. \ac{CMT} is a combination of
\ac{CMP} and \ac{FG-MT}. With \ac{FG-MT}, a processor core is capable
of quickly switching between active threads. As a result, memory
latency of a thread can be filled by executing another active thread.
In this way, the processor core tries to do active work in each cycle.
A SPARC T3 processor has up to sixteen cores, each supporting eight
hardware threads per core. In our case, a SPARC T3-4 server is used.
This server is shown schematically in Figure \ref{fig:t3:overview}. As
the name suggests, the SPARC T3-4 has four SPARC T3 processors running
at 1.65GHz. This gives a total of 64 T3 cores and a total number of
512 hardware threads.

Within a T3 core there are two execution units, one per four hardware
threads. Additionally, a single floating point / graphics unit (FGU)
is present for all the eight hardware threads. Level 1 caches
are present in each individual core. For instructions a 16KB
instruction cache is available and data can be stored in a 8KB data
cache. The level 2 cache are shared between all the cores on a single
SPARC T3 processor using a crossbar switch. The cores in the SPARC T3
processor are kept consistent using coherency units. Finally, each T3
core has its own memory management unit for virtual memory management.
For the instruction data, a instruction \ac{TLB} of 64 entries is
present and the data \ac{TLB} has 128 entries.

For comparing the performance of the scenario-based DSE on the SPARC
T3-4, we have also ran the original implementation of our
scenario-based DSE on a Sun Fire X4440 \cite{oracle4440} compute
server with four quad-core AMD Opteron 8356 processors running at
2.3GHz. In contrast to the SPARC T3-4, this machine runs CentOS Linux.
   
\section{Unoptimized implementation}\label{sec:impl}
\begin{figure*}[!t] \centering
\begin{subfigure}[b]{3.2in} \centering
\includegraphics[width=3.2in]{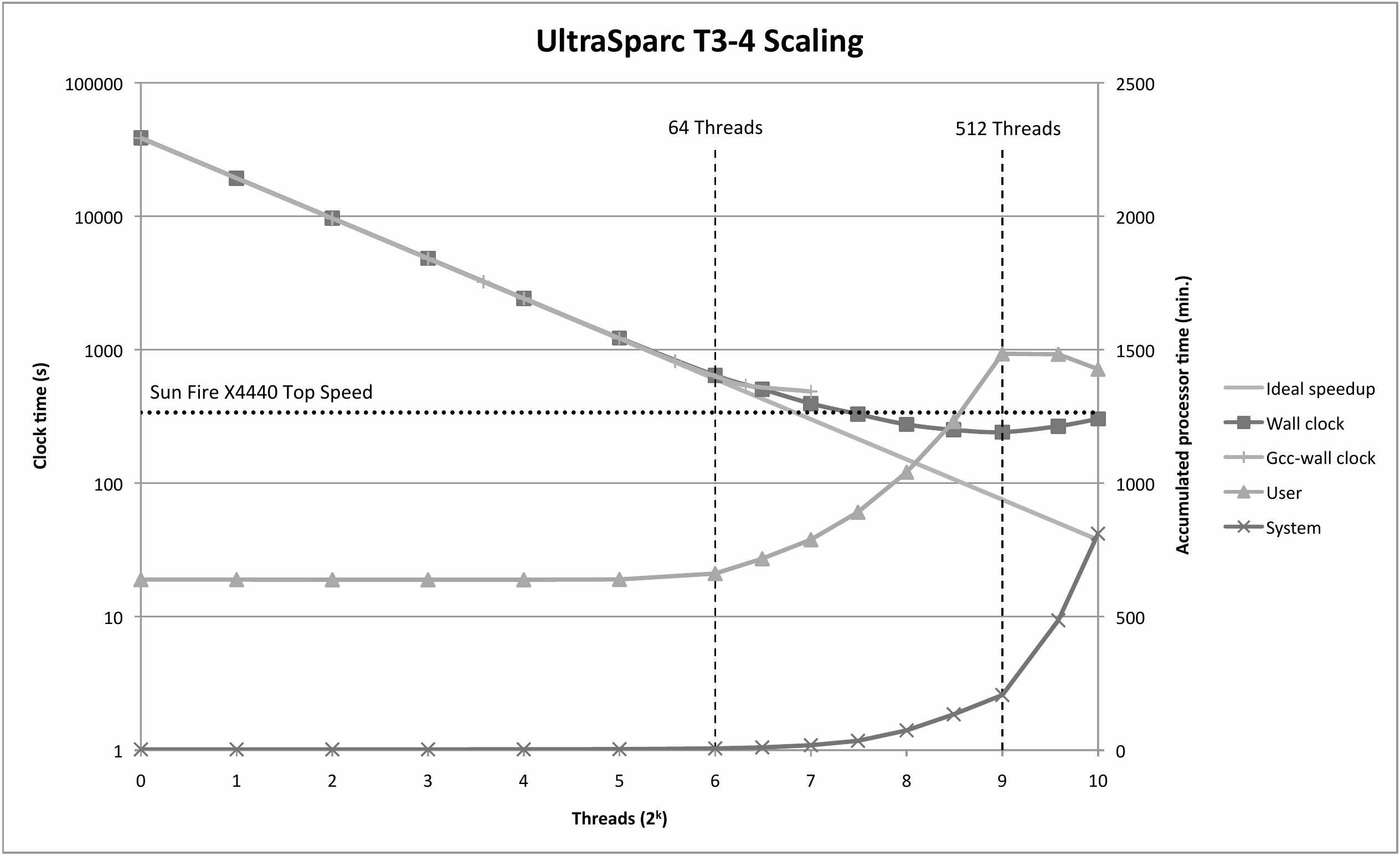} \caption{32 Bits.}
\label{fig:t3:plain32} \end{subfigure} \hfill \centering
\begin{subfigure}[b]{3.2in} \centering
\includegraphics[width=3.2in]{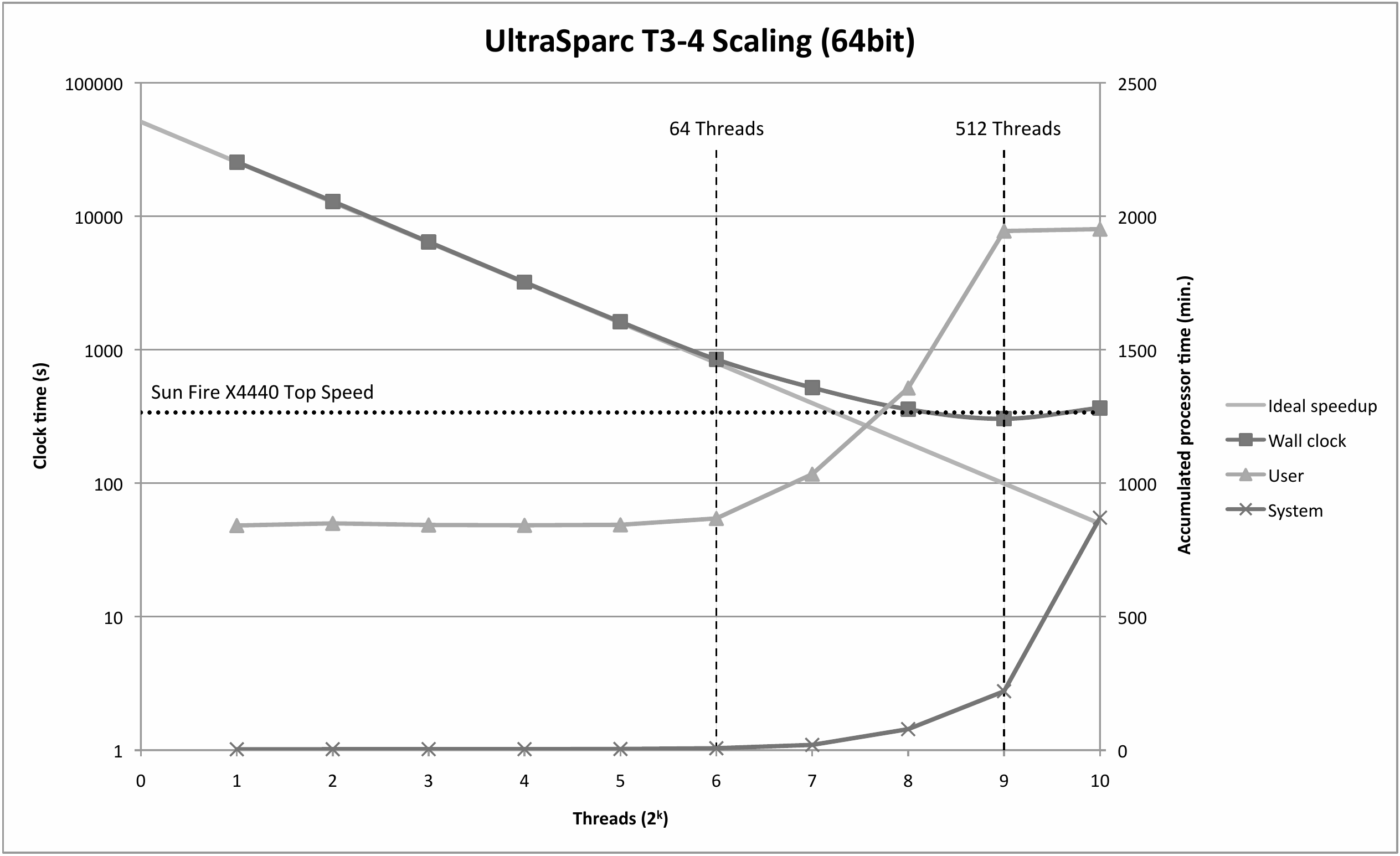} \caption{64 Bits.}
\label{fig:t3:plain64} \end{subfigure}

\caption{The performance scaling of Sesame on the Oracle SPARC T3-4
for the initial 32 and the 64 bit compilation.}
\label{fig:t3:plain}
\end{figure*}

At first, we have ported the scenario-based DSE to the SPARC T3-4
without any optimization. As the SPARC T3 processor supports 64 bits,
we had the possibility to compile for both $32$ and $64$ bits. In order to
see which option was best, we compared the performance scaling of both
options.

In Figure \ref{fig:t3:plain32}, the performance scaling is shown with
the unmodified scenario-based DSE and default compilation
flags\footnote{Optimization \texttt{-O3} for gcc and \texttt{-xO5} for
the Sun cc compiler. Other flags are tried, but do not have any
significant effect}. During this experiment, the size of the workload remains fixed and the
number of worker threads is varied. At first,
we investigated the total wall clock time of the
application. The logarithmic horizontal axis shows the number of
threads that are used to simultaneously process the simulations, whereas the left
vertical axis shows the wall clock time, also using a logarithmic
scale. As a consequence, the ideal speedup manifests itself with a
straight line.

With the 32 bit compilation, a comparison is made between the compiled
code of gcc and compiled code of the Sun cc compiler. Up to 128
threads, the performance of the Sun cc compiler and the gcc compiler is
similar and, therefore, we have decided to use the Sun cc
compiler for all the experiments. For this compiler, the speedup
of the code is completely linear until 64 threads. With more than 64
threads the speedup quickly decreases. At the optimal point of 512
threads, the Oracle SPARC T3-4 is 29 percent fast than our Sun Fire
X4440 server.

In order to identify the cause of the decrease in speedup, the user
and system time is also plotted in Figure \ref{fig:t3:plain}. For this
graph, the right vertical axis shows the accumulated processor time
in minutes. Since the amount of work is constant, the expected
behavior is that the accumulated user time remains the same. This is
indeed the case when the number of threads is between $1$ and $64$. After
this point, however, the user time starts to increase. The same is true
for the system time: 
due to the increased number of threads, the complexity of
the coordinating tasks of the operating system starts to increase.
This can be clearly seen when the system is overloaded with more than
512 threads.

For the user time, however, the increase is not as you would expect.
The most plausible explanation is the increase is related to the number of
physical cores. There are 64 physical cores on the Oracle SPARC T3-4.
When the number of threads is lower or equal than the number of
physical cores, the speedup is linear. With more than 64 threads the
hardware threads should lead to a further improvement of the
performance. Still, this improvement is not linear anymore. Although
hardware threads have their own execution units, a small part of the
functionality is shared with
the other hardware threads on the physical core. In order to identify
this shared functionality in our program and to minimize these dependencies
between the tasks, we have applied profiling. This will be described
in the following two sections.

Finally, the comparison between the 32 bits (Figure
\ref{fig:t3:plain32}) and the 64 bit implementation (Figure
\ref{fig:t3:plain64}) shows that the 32 bit option on the SPARC is
superior to the 64 bit option. In our reference system, the x86 based
Sun Fire X4440, 64 bits compilation is done, which is clearly faster
than when 32 bit compilation is used. In the SPARC T3-4, however, 32
bit compilation is better. In Figure \ref{fig:t3:plain64} the
performance scaling of the 64 bit version of our application can be
seen for the Oracle SPARC T3-4. The graph shows exactly the same trend
as the 32 bit compilation (Figure \ref{fig:t3:plain32}). However, the
absolute performance is worse than the 32 bit compilation.  The
difference between our x86 based reference system and the SPARC T3-4
is that in the x64 architecture 64 bit compilation enables some
additional architectural features. For the SPARCV9 architecture,
however, these optimizations are already available in the 32 bit
compilation. As a result, on the Oracle SPARC T3-4 scenario-based DSE
only suffers from the increased memory footprint and does not benefit
from additional architectural features.

\section{Profiling of a Single Sesame Call}\label{sec:profSesame}
\begin{figure*}[!t]
\centering
\includegraphics[width=\textwidth]{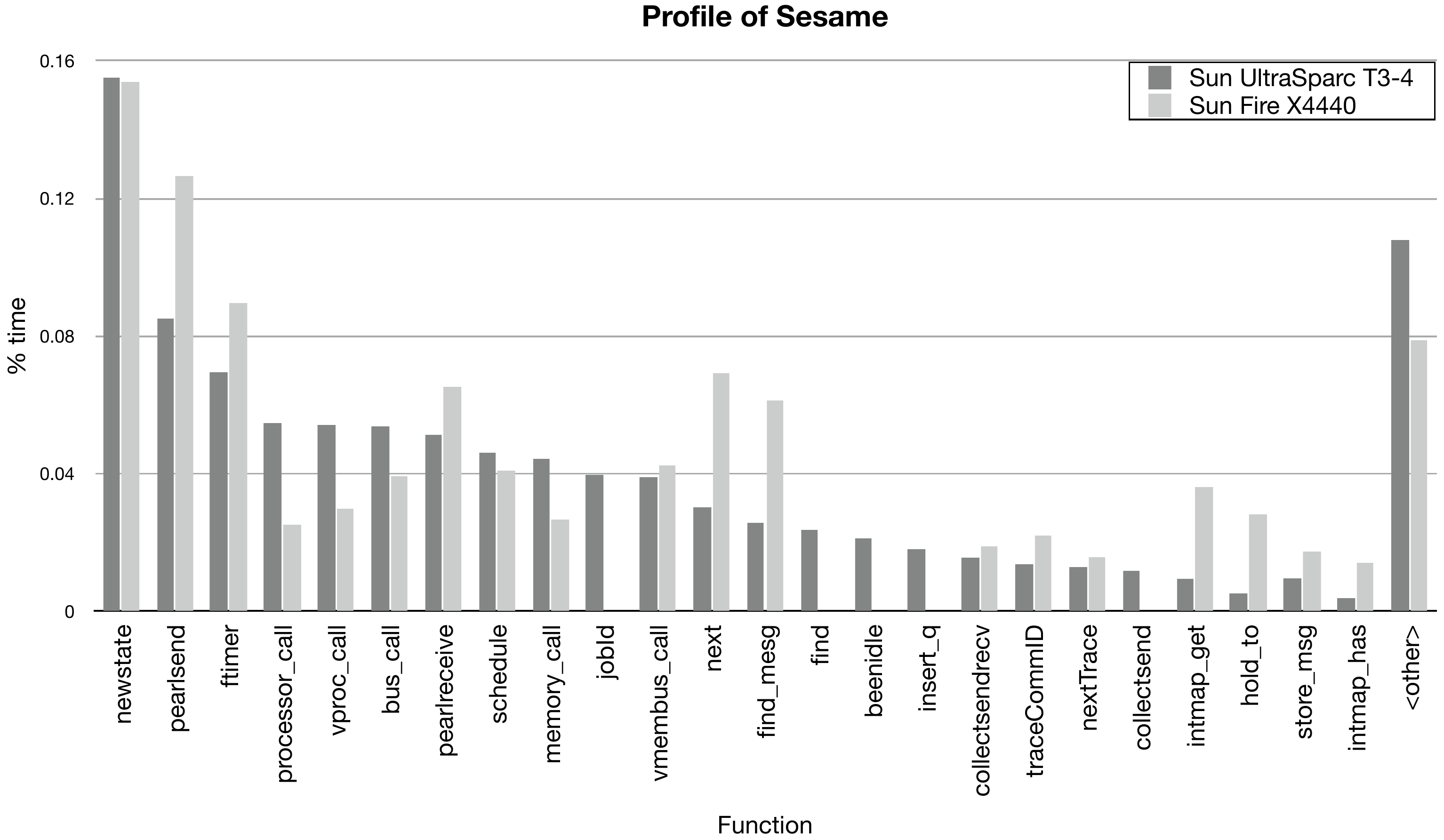}
\caption{The function profile of a single Sesame simulation obtained with gprof.}
\label{fig:t3:gprof}
\end{figure*}

We start by profiling an individual Sesame simulation on the Oracle SPARC T3-4 and the Sun Fire
X4440. As the Sun Fire X4440 is running CentOS linux, we use gprof
\cite{graham04} to obtain the profile. In Figure \ref{fig:t3:gprof}
the profile with the functions that use the largest amount of
exclusive user time is shown.
For the performance of a single Sesame call, there is a large
difference between the Sun Fire X4400 and the Oracle SPARC T3-4. The
single thread performance is approximately 7 times slower than the Sun
Fire X4440.  It is to be expected that the single thread performance
is worse on the Oracle SPARC T3-4, than it is on the Sun Fire X4440.
The individual core of the SPARC T3-4 is not only simpler, but the clock frequency is
also lower. This difference does not seems to be related to
differences in the function profile. As there are no significant
differences in the function profile, the porting of the scenario-based
DSE to the SPARC T3-4 does not introduce new bottlenecks in a single
Sesame simulation. 
Therefore, we decided that there
is no need to invest time in the optimization of a single simulation.
Rather, there is focused on optimizing the scenario-based DSE that
executes multiple simulations simultaneously.

\section{Profiling of the Scenario-based DSE}\label{sec:profSDSE}
\begin{figure*}[!t]
\centering
\begin{subfigure}[b]{2.8in}
\centering
\includegraphics[width=2.8in]{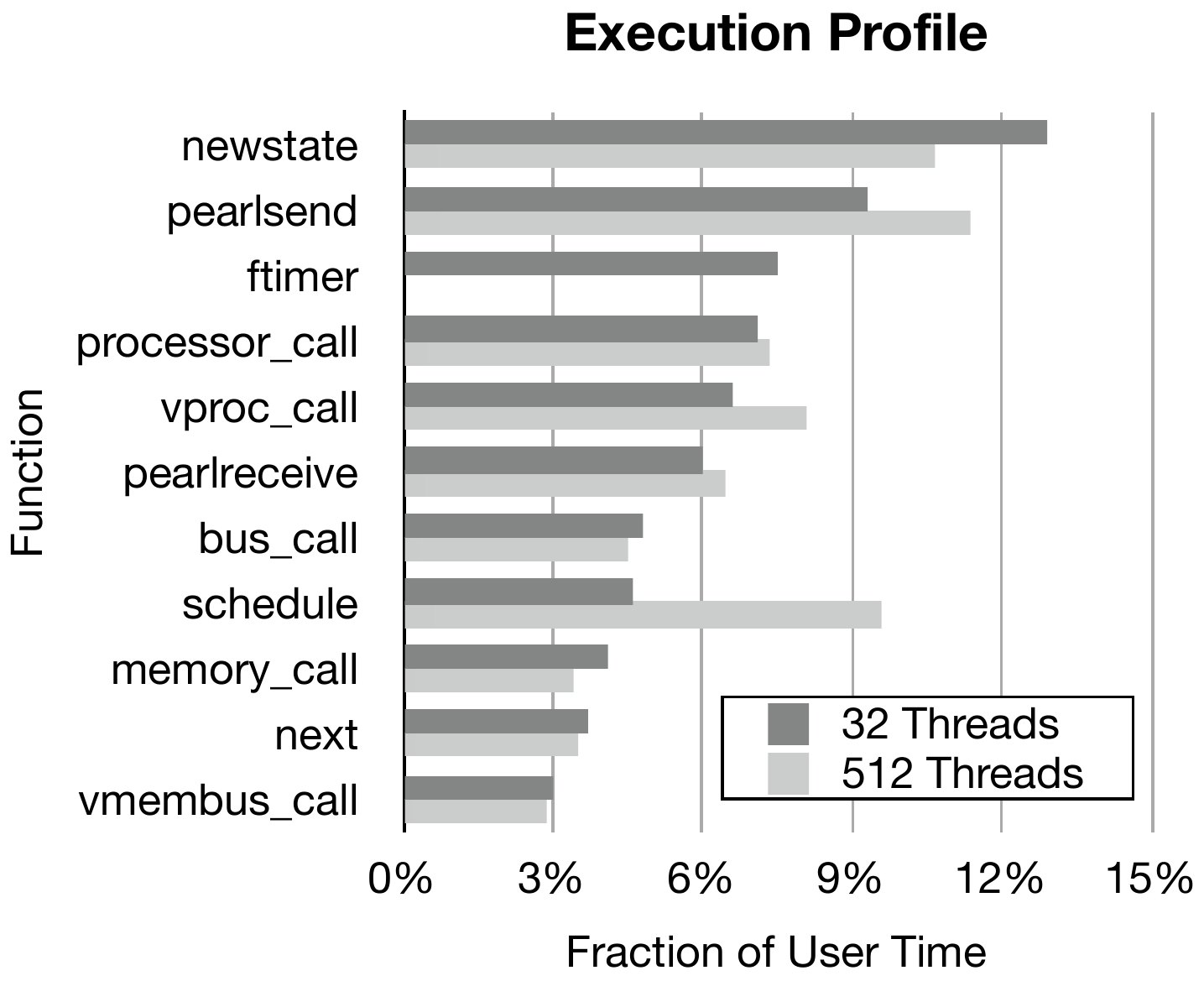}
\caption{Profile}
\label{fig:t3:prof}
\end{subfigure}
\hfill
\centering
\begin{subfigure}[b]{2.8in}
\centering
\includegraphics[width=2.8in]{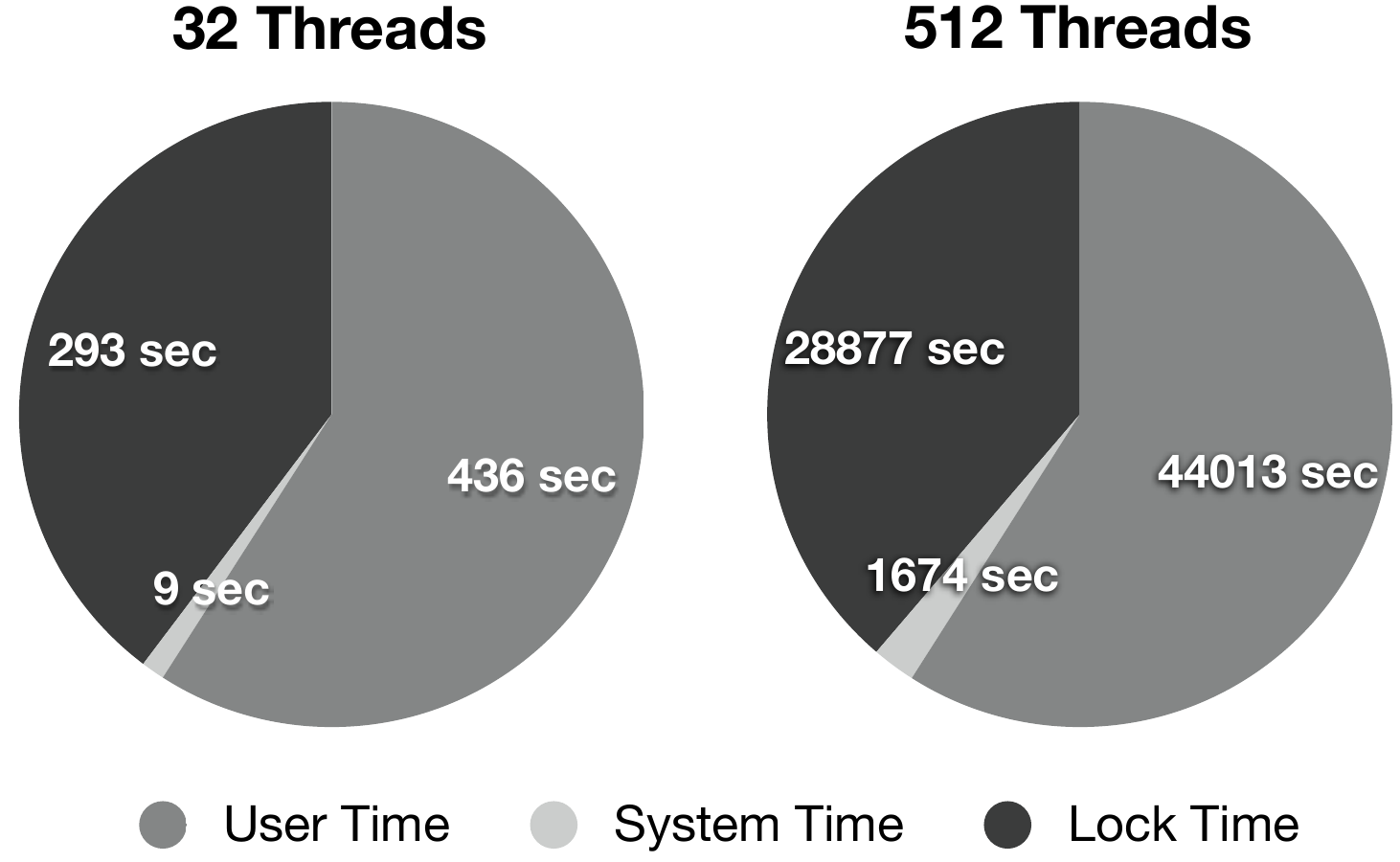}
\caption{Breakdown}
\label{fig:t3:breakdown}
\end{subfigure}

\caption{The profile and execution time breakdown (accumulated over
all threads) of the scenario-based DSE with $32$ and $512$
simultaneously running worker threads. In contrast to the 32 thread
experiment, the $512$ thread experiment uses full function inlining.}
\label{fig:t3:sunprof}
\end{figure*}

Next, the complete scenario-based \ac{DSE} framework is profiled. For
this purpose, the Oracle Solaris Studio Performance Analyzer \cite{grove10} is used. Figure
\ref{fig:t3:prof} shows the function profile of a scenario-based DSE
where $32$ worker threads are running simultaneously. The profile shows the
same behavior as the profile of a single Sesame simulation.
However, during the Sesame simulation there are some clear hot spots. One of
these hot spots is \verb#ftimer#. This function only returns (after a
short calculation) the current simulation time. To improve the
performance, we have enabled function inlining in the Sun cc compiler for functions
from external libraries (\verb#ftimer# is a function in a dynamic
shared library that is normally imported at runtime). The
effect of inlining is clearly visible in the profile of Figure
\ref{fig:t3:prof}, where the \verb#ftimer# function is inlined with
the $512$ thread experiment. As the function \verb#ftimer# is not present
any more, its user time drops to zero. For some functions (e.g., \verb#pearlsend# and \verb#schedule#), however,
the relative user time increases significantly.
Due to the heavy use of inlined
functions, the relative amount of computation time increases and, as a
result, the fraction of the total exclusive user time that is spent in
these functions.

In Figure \ref{fig:t3:plain} we noticed that there was an
increase in user and system time for situations where the number of
threads was larger than $64$. To analyze this growth, the breakdown
of the total processing time accumulated over all threads is shown in
Figure \ref{fig:t3:breakdown}. Apart from the absolute values, the
fractions of user, user lock and system time are more or less equal.
Still, the overhead of the scenario-based DSE seems to be unreasonably
large. Although none of the scenario-based DSE framework function show
up in the profile (Figure \ref{fig:t3:prof}), the amount of user lock
time is significant. Most likely, part of this lock time is due to the
lock-based job queue of the scenario-based \ac{DSE}. To resolve
this issue, we have redesigned our scenario-based DSE to use a
lockless job queue. This design will be described in the next section.

\section{Final design}\label{sec:final}
The original design of the workpool of the scenario-based DSE used a
queue based on mutexes and condition variables to enforce that
only one thread at the time retrieves a job from the job queue. Such a lock-based
design works satisfactory if the number of worker threads are low.
However, when there is scaled up to 512 threads the contention becomes relatively
large. This is substantiated by the amount of lock time in the time breakdown in
Figure \ref{fig:t3:breakdown}.

\begin{figure}[t]
\centering
\includegraphics[width=2in]{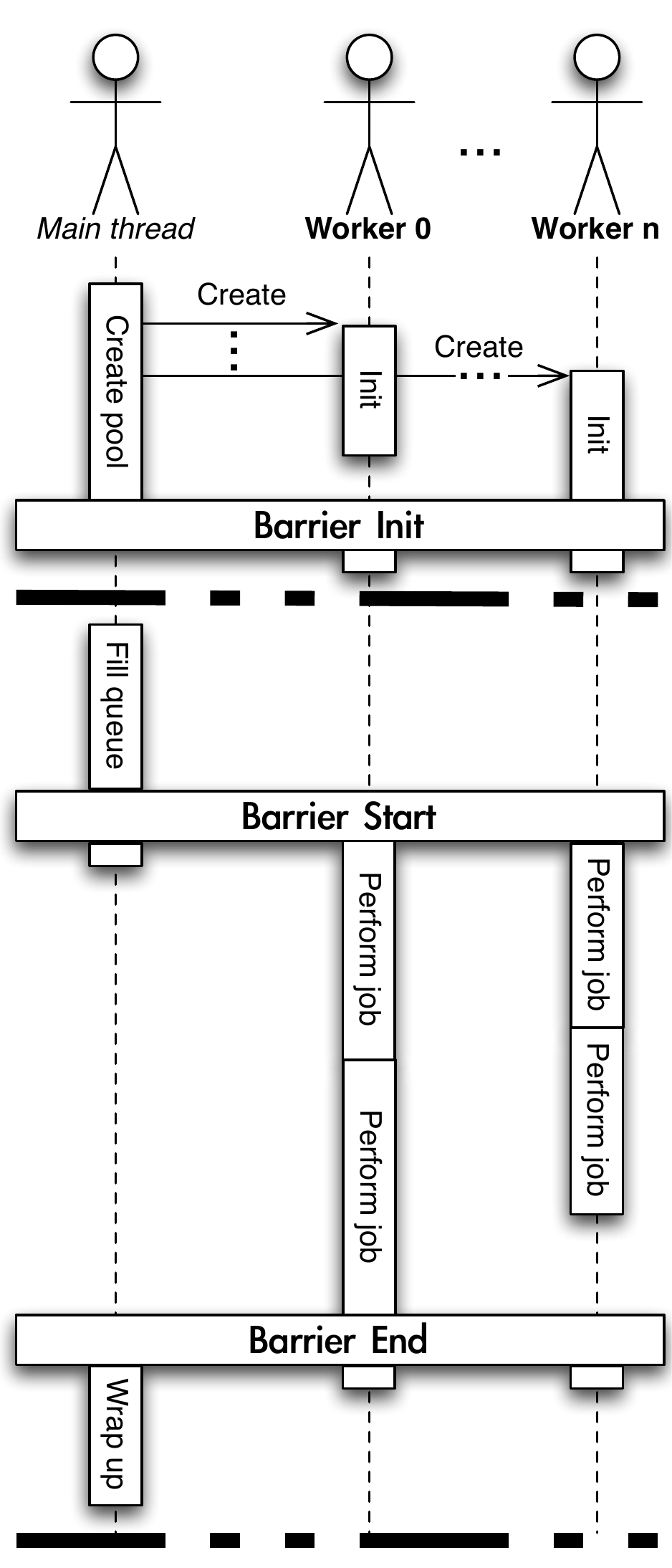}
\caption{The "lockless" implementation of the scenario-based DSE work pool with 3 worker threads}
\label{fig:t3:design}
\end{figure}

Figure \ref{fig:t3:design} shows the lockless design
of the workpool. The lockless implementation is largely based on
volatile variables and atomic operations.  Additionally, synchronization is achieved
using a barrier. During execution, two stages can be
distinguished:
initialization and the main execution. Initialization is triggered by the main thread. The main thread will
create all the worker threads one by one and wait
until the worker threads are initialized. Next, the worker threads
will initialize their data structure. After the worker threads are all
ready for execution, the threads are synchronized using a barrier
(\textsc{Barrier Init}).
 
Within the main execution, the job queue is filled and processed by
the worker threads. During the filling of the queue, all the worker
threads are blocked on a barrier (\textsc{Barrier Start}). The filling
of the queue is done by the main thread. This involves the allocation
of a vector with job descriptions and, next to this, the atomic variables
are initialized. There are two atomic variables: 1) the \verb#cur# pointer that
refers to the first unprocessed job and 2) the \verb#end# pointer that
refers to the last job. After initializing the queue, the main thread will also 
synchronize on the start barrier.

When all threads are started the jobs will be processed. Each of the
jobs will be handled by a single worker threads that will start a 
Sesame simulation in an external process using the
\verb#system()# command\footnote{Unfortunately, it is not possible to
easily integrate Sesame in the evaluator due to the large number of global
variables in the program}. In the meanwhile, the main thread is
blocked on the final barrier (\textsc{Barrier End}) until the complete
job queue is handled. To fetch a job, a worker thread atomically
increments the \verb#cur# pointer and obtains the current value. In
case the value is smaller or equal to the \verb#end# pointer, the
specific job will be fetched from the queue. Otherwise, the worker
thread will also synchronize on the final barrier.

Once all the threads have reached the end barrier the main thread will
wrap up. This involves the destruction of the queue and make it ready
for the next batch of evaluations.  In the meanwhile, the worker
threads are already waiting on the start barrier. This design allows
us to process multiple batches without recreating the workpool for
each generation in the scenario-based DSE.

\section{Experiments}\label{sec:exp}
Until now, we have described the port of the scenario-based DSE to the
SPARC T3-4. The focus of this porting procedure was to run the
evaluation of a batch of simulations as fast as possible. In this
section we will present the final results using three types of
experiments. 
During the experiments, the final design of the scenario-based DSE is
used with a fixed workload that consists of $1000$ individual
simulation jobs.
The first two experiments will analyze the influence of
the type of heap allocation and the type of scheduling. Next, we will
give a short remark on the wall clock time accuracy of the
Oracle Solaris Studio Performance Analyzer.
This is followed by an experiment that explains the increase of user time
with an increasing number of worker threads. Finally, we can show with a
final experiment the scalability of the scenario-based DSE.

\subsection{Heap Allocation}
\begin{figure}[!t]
\centering
\includegraphics[width=0.45\textwidth]{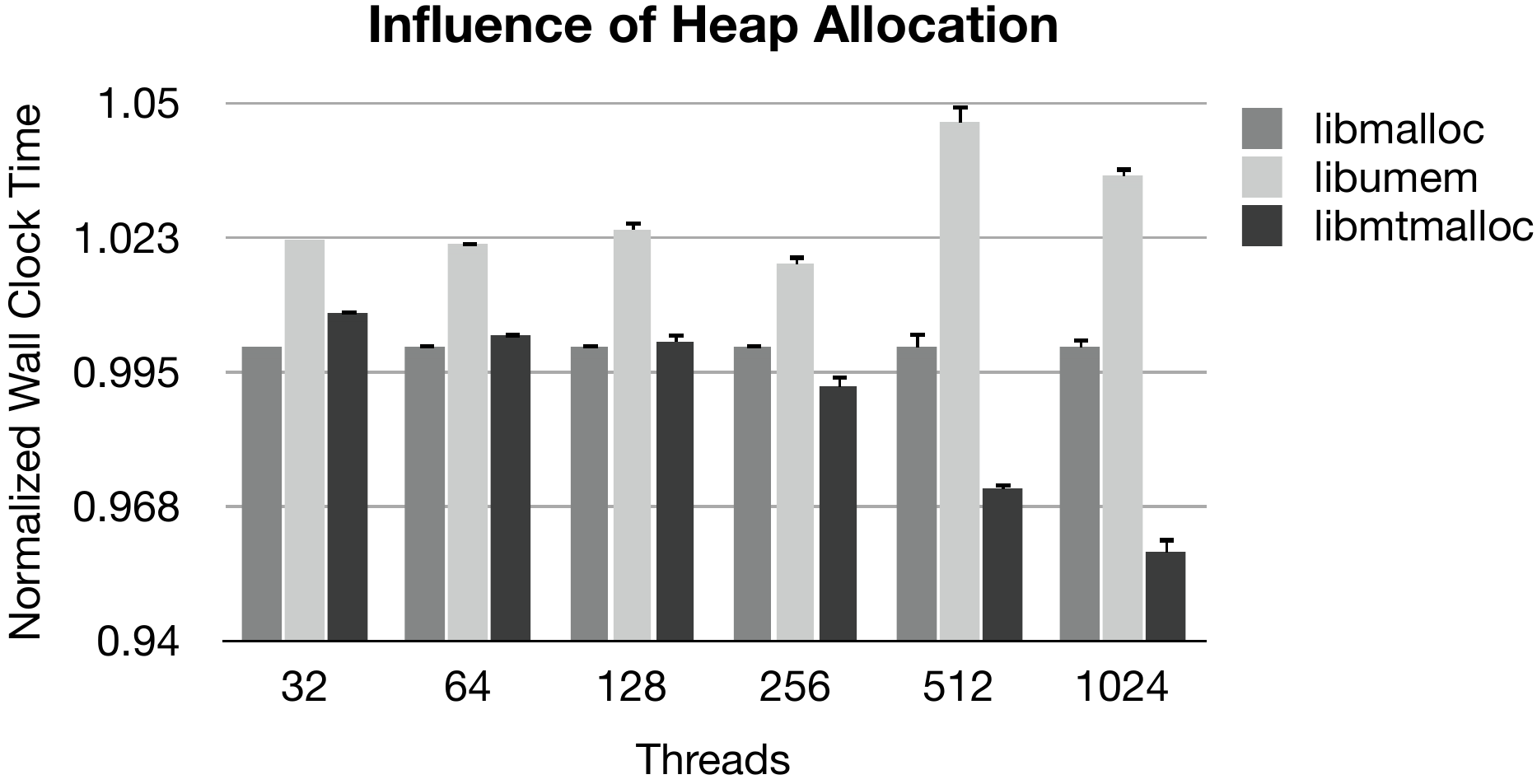}
\caption{The influence of the different types of heap allocation on
the execution time of the scenario-based DSE.}
\label{fig:t3:heap}
\end{figure}

During the profiling with the Oracle Solaris Studio Performance
Analyzer, one hot spot with respect to system time was the function
\verb#take_deferred_signal()#. When digging deeper in the function
stack, we found out that this function becomes hot due to mutex locks
in \verb#malloc# and \verb#free#. The default malloc library on Oracle
Solaris uses a single heap for all the different threads.
During system calls like \verb#malloc# and \verb#free#, the access to
the shared heap is guarded by mutex locks. In a system where there are many hardware threads,
such as the SPARC T3-4, this quickly can become a bottleneck of an application.

\begin{figure*}[!t] 
\centering          
\begin{subfigure}[b]{2.75in}
\centering
\includegraphics[width=2.75in]{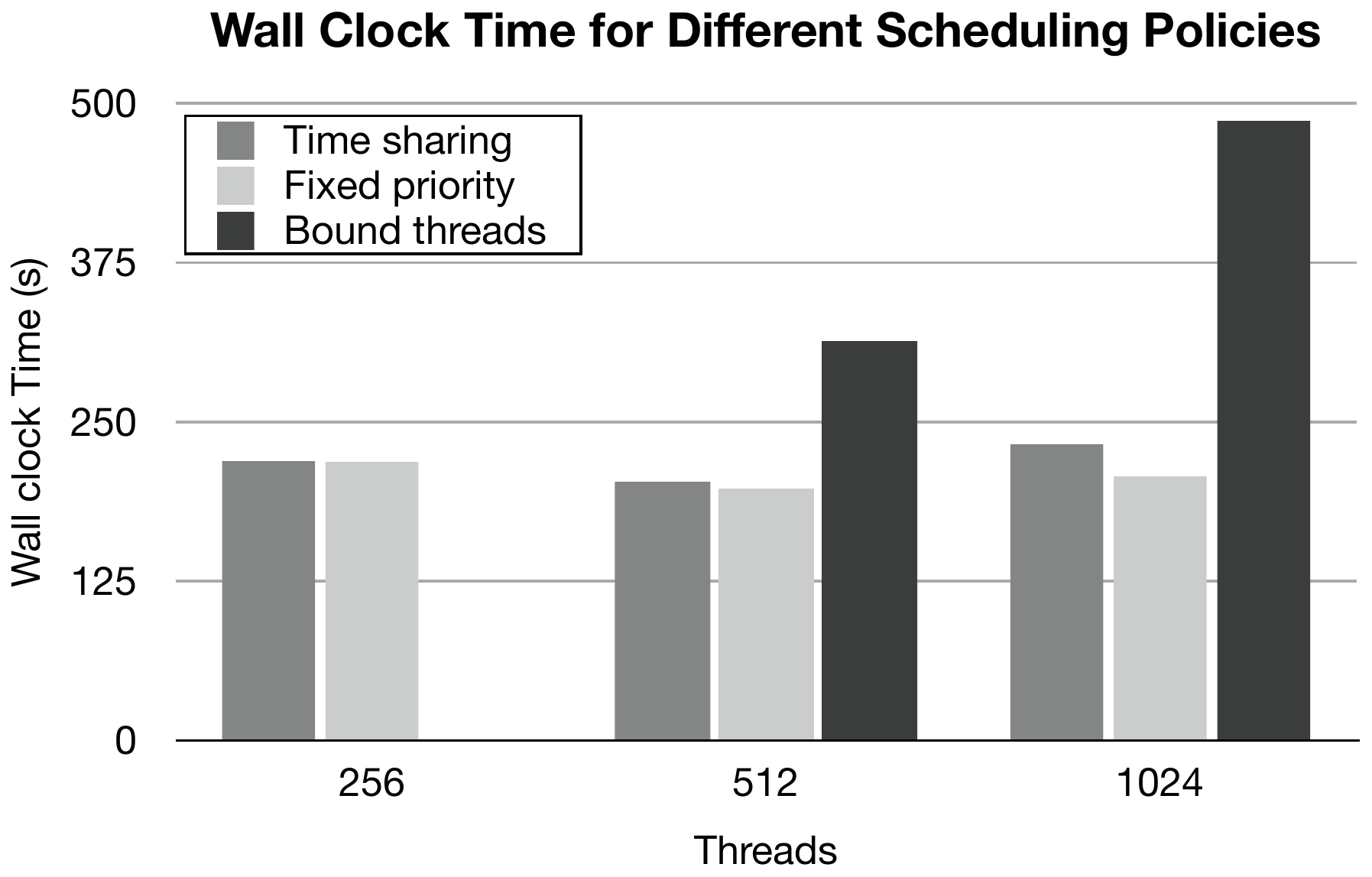}
\caption{Execution Time}
\label{fig:t3:sched-time}
\end{subfigure}
\hfill
\centering
\begin{subfigure}[b]{2.75in}
\centering
\includegraphics[width=2.75in]{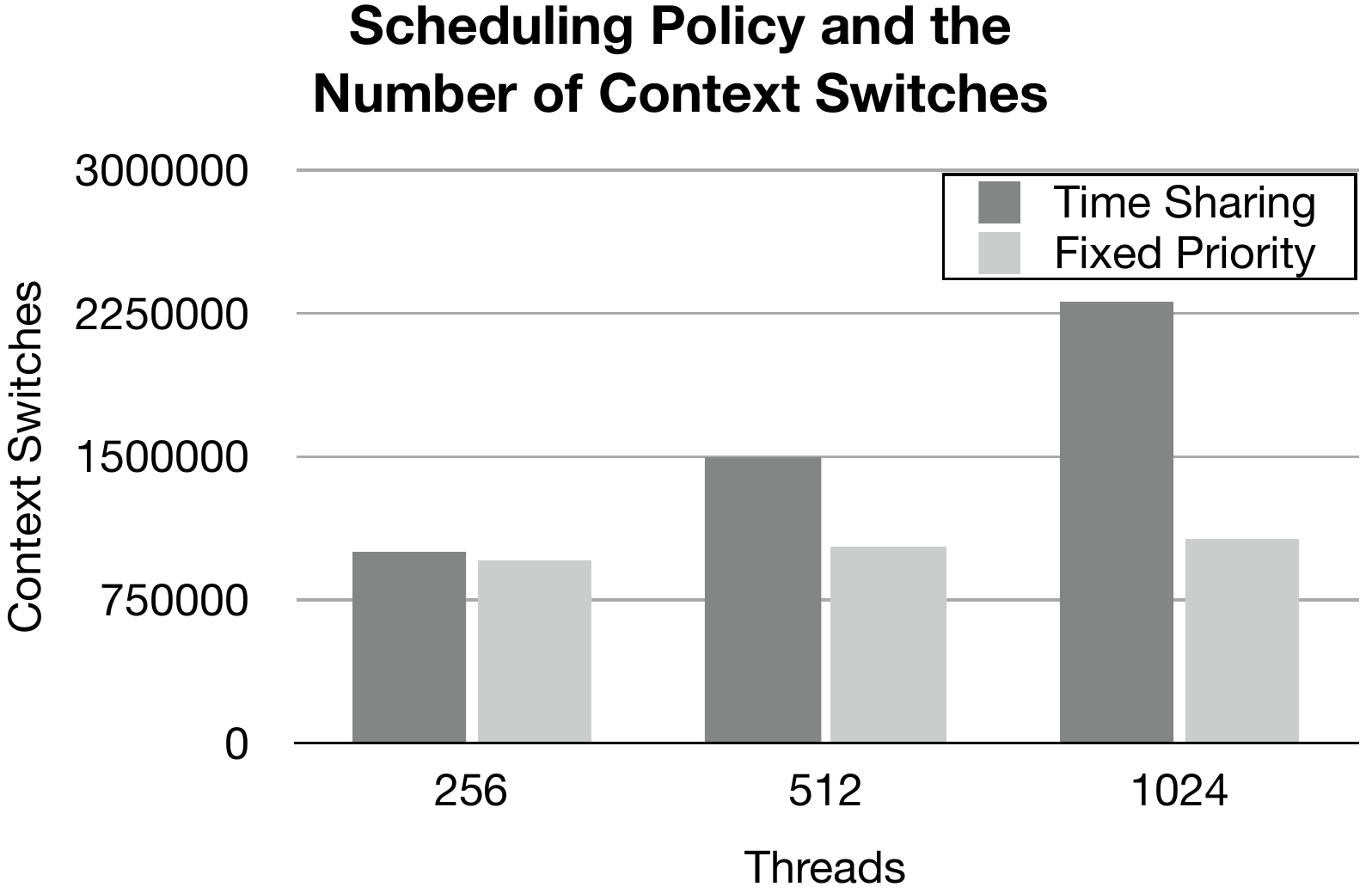}
\caption{Context Switches}
\label{fig:t3:sched-switch}
\end{subfigure}

\caption{The wall clock execution time with different scheduling
policies and the relation to the number of context switches.}
\label{fig:t3:sched}
\end{figure*}

Fortunately, Oracle Solaris provides more heap allocation strategies. Next
to the default strategy, a multithreaded malloc (\verb#libmtmalloc#)
and a type of slab allocator (\verb#libumem#). We compared the
performance of our scenario-based DSE framework on these three types of heap
allocation. The experiment is performed for a variable number of
worker threads, where each experiment is repeated six times.

The result of the experiment is shown in Figure \ref{fig:t3:heap}.
In the results the wall clock time of every experiment is normalized to the
average wall clock time of the default heap allocation scheme and the
error bars show the standard error of the mean. At first, it is clear
that our application is not suited for slab allocation. In the case
of slab allocation, the heap allocator tries to reduce the memory
fragmentation by preallocating memory slots of a certain type. When
these types are allocated frequently, this quickly provides allocated
memory with hardly any fragmentation. This approach may be well suited
to kernel objects (In fact, \verb#libumem# is a user space
implementation of the original slab allocator inside the kernel), but
in our scenario-based DSE framework \verb#libumem# is significantly slower for all 
cases. It also does not solve our bottleneck problem as it has the
same mutex locks as the default heap allocator.

The multithreaded heap allocator \verb#libmtmalloc#, however, has
split the heap into individual segments for each separate thread.
This requires more heap space, but locally the dynamically allocated data
can be created concurrently for each of the different threads without
using locks. Our
results show that for situations where a relative modest number of
threads are used, \verb#libmtmalloc# is slower. In this case we only
suffer from a larger heap space. Increasing the number of workers,
\verb#libmtmalloc# is becoming faster than the default heap allocator.
The more worker threads there are, the more lock contention is present
in the default heap allocator. This lock contention is not present in
the multithreaded heap allocator, what is especially visible when we
overload the system with $1024$ worker threads.

\subsection{Scheduling}

Another aspect that can influence the performance is the scheduling policy of
the process. Solaris 10 allows us to set the scheduling class of a
process with the command \verb#priocntl#. For a normal user, there are
two possible classes: 1) time sharing and 2) fixed priority. Time sharing
periodically recalculates the priority of a process to give each
process an equal part of the processing time, whereas in the case of
the fixed priority it remains equal for the total lifetime of the
process.

As in our scenario-based DSE framework separate processes are used for each
individual simulation, the scheduling can affect the
performance. As shown in Figure
\ref{fig:t3:sched-time}, the desired behavior of the performance is
that it improves until all the 512 hardware threads of the SPARC T3-4
are utilized. After this point, the performance should degrade very
slowly. This is the case with the fixed priority policy. For time
sharing, however, the performance of the scenario based \ac{DSE} framework degrades faster and using 1024
threads it is even slower than using 256 worker threads.

Most likely, the reason for the degraded performance of the time
sharing policy is the number of context switches. In order to quantify
the influence of context switches, we have used the standard C library
function \verb#getrusage#. The results in Figure
\ref{fig:t3:sched-switch}, show indeed a correlation between the
number of context switches and the degraded performance. For the fixed
priority policy the number of context switches remains constant with
an increased number of worker threads. The number of context switches
for the time sharing policy, on the other hand, increase
simultaneously with the number of worker threads.

We realized that it should be beneficial that each worker process
keeps the affinity with the T3 core where it is running. In this way,
a more efficient cache usage can be achieved. To achieve the (virtual)
processor affinity, all the worker threads are bind to one of the
hardware threads with the system call \verb#processor_bind#. Since the
manual did not provide us with a clear mapping of the processor
identifier and the hardware thread, we only performed this experiment
for 512 and 1024 worker threads. In this case, it is relatively easy to spread to
workers over the architecture. Looking to the performance in Figure
\ref{fig:t3:sched-time}, our simple worker mapping scheme does not give
satisfactory results. Undoubtedly, better worker mapping schemes can
be identified, but we do not expect to obtain significant improvements.

\subsection{User Time Scaling}\label{sec:tlb}
\begin{figure*}[!t]
\centering
\includegraphics[width=4.5in]{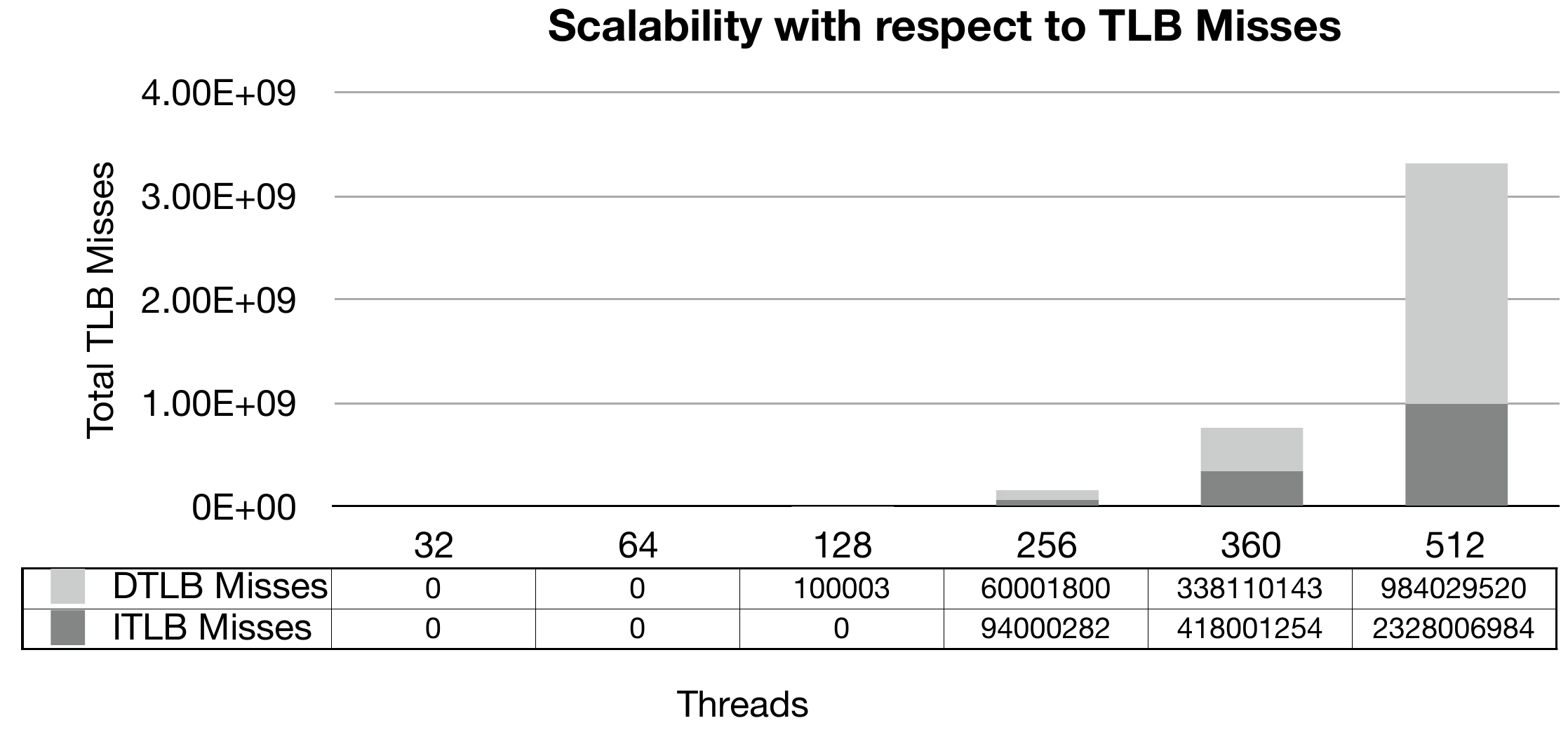}
\caption{The number of TLB misses in relation to the number of worker threads}
\label{fig:t3:tlb}
\end{figure*}

A much larger potential improvement can be gained if we resolve the
increase in user time. Going back to the scalability graph in Figure
\ref{fig:t3:plain}, we identified that once the number of worker
threads is larger than 64, the total user time of the application
starts to increase. However, as the workload remains constant, the
user time should remain constant irrespective of the number of worker
threads. If we can reduce the increase in user time, the total speedup of the
application (compared to sequential execution) can be improved
significantly.

In order to identify the cause of the increase, a significant amount of research with
the Oracle Solaris Studio Performance Analyzer was required. The
hardware counters finally gave us the solution. In Figure
\ref{fig:t3:tlb}, the number of \ac{TLB} misses can be seen for the
instruction memory and the data memory. Obviously, the number of \ac{TLB} misses
show a high correlation between the increase in user time. Until 64
worker threads, the user time was constant: the \ac{TLB} experiment
shows us that in this case there are no \ac{TLB} misses. After this
point, the user time is increasing and this is reflected by a
skyrocketing number of \ac{TLB} misses. When a \ac{TLB} miss occurs in
user mode, it is also resolved in user mode. So, the time to resolve
the \ac{TLB} misses is also added to the user time. With more than 3.3
billion \ac{TLB} misses, it is to be expected that a large increase in
user time is observed.

The architecture of the SPARC T3-4 also explains why the threshold is
at 64 worker threads. Each worker thread uses separate processes to
perform the simulations. As a result, each worker thread needs his own
private entries in the instruction and data \ac{TLB}. Until 64 worker
threads, each worker thread can run at a separate T3 core and have his
own \ac{TLB}. However, when the number of worker threads is larger
than the number of T3 cores, the \acp{TLB} are shared. In the case
that all 512 hardware threads are utilized, each hardware thread can
have only 8 entries in the instruction \ac{TLB} and 16 entries in the
data \ac{TLB} ($\frac{1}{8}$ of the entries in the shared \ac{TLB} of
the T3 core).

It is hard to resolve the \ac{TLB} misses. At first, Sesame includes a
large number of shared libraries. Secondly, a large amount of data is
used during the simulation. This involves input data like the workload
of the embedded system, the description of the application and the
architecture of the embedded system. Additionally, a large amount of
output data is produced that is temporarily stored in memory. 

\begin{figure}[!b]
\centering
\includegraphics[width=2in]{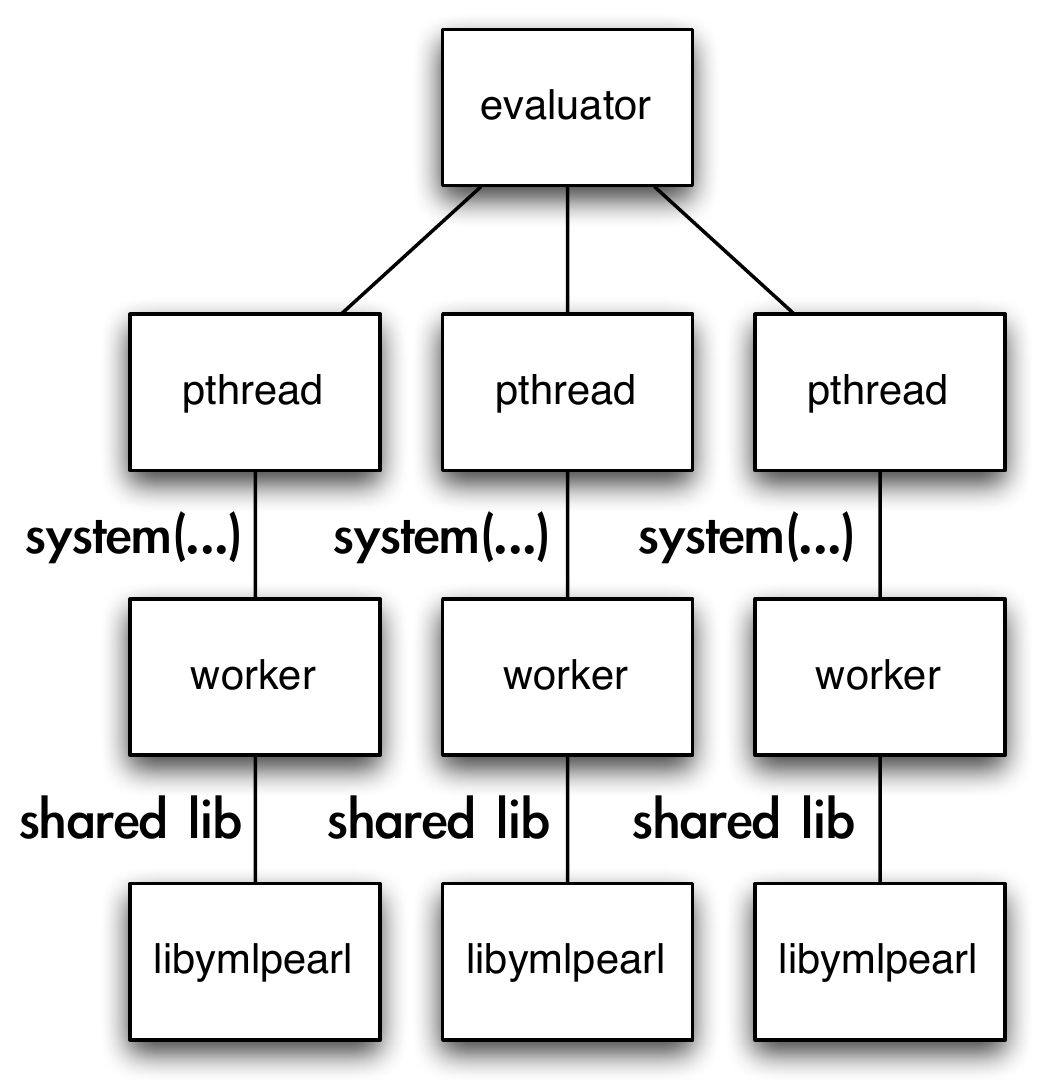}
\caption{The relation between threads and processes in the scenario-based DSE}
\label{fig:t3:rel}
\end{figure}

The relation between threads and processes, as depicted in Figure
\ref{fig:t3:rel}, explains the TLB problem. For the scenario-based \ac{DSE} the first two levels of the
tree are running in a single process (the evaluator and worker threads).
Sesame workers, however, are separate processes. The Oracle Solaris Studio
Performance Analyzer learns us that all the \ac{TLB} misses occur in a
large shared library named \verb#libymlpearl#.

A potential improvement is to incorporate Sesame in the evaluator as a
shared library. In this way the simulation is a function call instead
of an externally running process. The big advantage in this case is
that \verb#libymlpearl# only needs to be loaded once. Shared data
between the simulation is in this situation shared between all the
workers.  However, from origin Sesame uses many global variables.
Thus, it would require a large (but absolutely not impossible)
implementation effort to turn Sesame into a loadable dynamic library. 

At the moment, the only possible improvement is to increase the page
size of the heap to 4MB instead of the default 8KB. This gives already
a performance improvement of more than four percent with respect to
the execution time with the 8KB page.

\subsection{Scalability}
\begin{figure}[!t]
\centering
\includegraphics[width=3in]{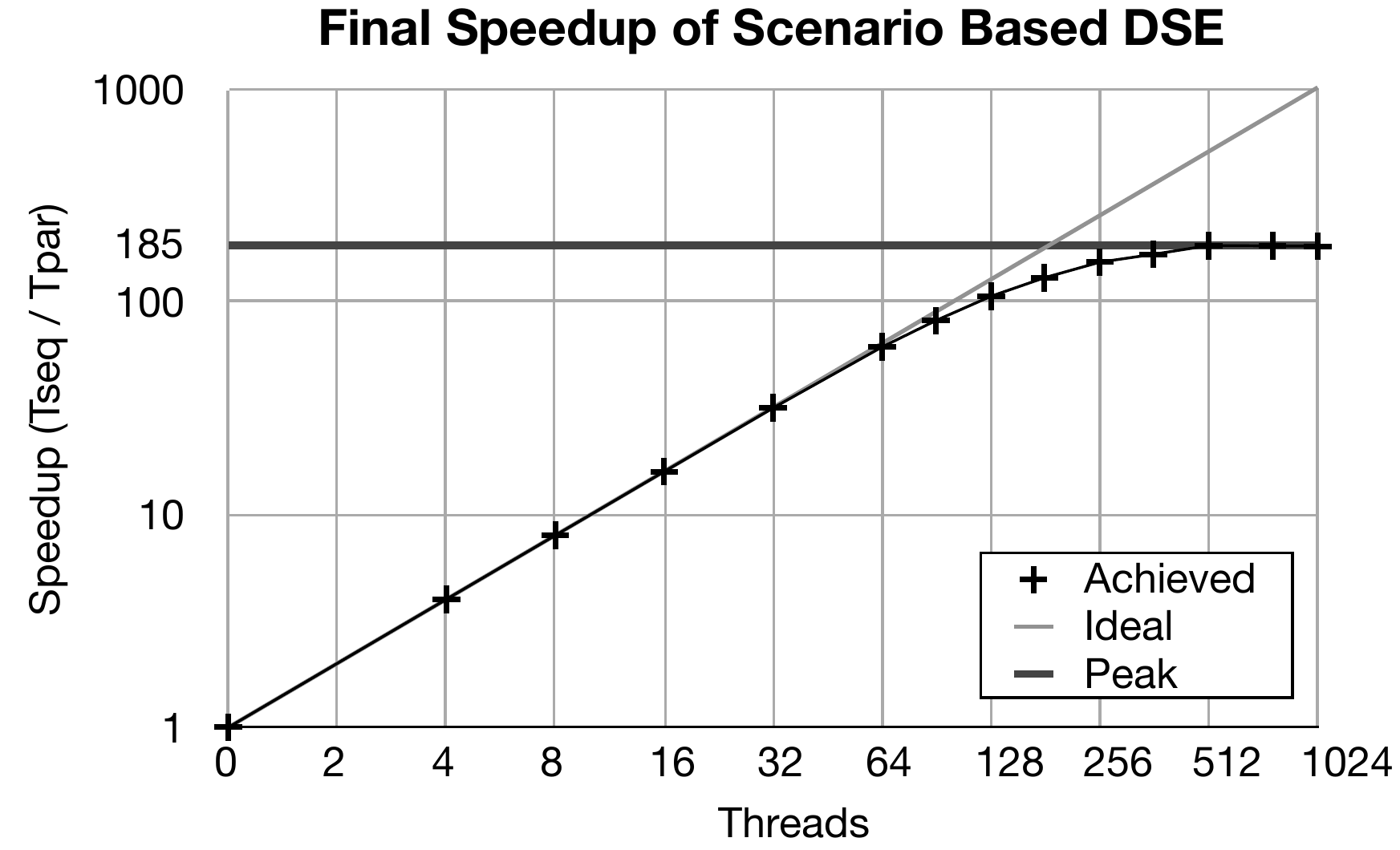}
\caption{The final scalability of scenario-based DSE on the SPARC T3-4}
\label{fig:t3:finalscal}
\end{figure}

With all these improvements, it is time to show the final scalability
of the application. For this experiment, we increased the size of the
workload to 10.000 jobs to be certain that the lack of sufficient
workload does not limit our speedup. The results are given in Figure
\ref{fig:t3:finalscal}. 

Linear speedup is achieved when the number of worker threads is less
or equal to the number of T3 cores. In this case, the chip
multiprocessing is exploited and most resources are private to the
worker threads. Examples of these resources are the level 1 caches,
the execution units and the \acp{TLB}.

For 128 threads the parallelized evaluator is 107 times as fast as the
sequential version. In this case, the average number of worker
threads per T3 core is two. Each worker thread has thus its own execution unit (as there
are two in each T3 core), but other resources like caches and
\acp{TLB} needs to be shared. As a consequence, the speedup is still
close to linear.

Above the 128 threads the execution units are
also shared between the worker threads. For these configurations, the
performance mostly suffers from the limited \ac{TLB} size. Hence, the maximal
obtained speedup is almost 185 times as
fast as the sequential execution. Given the fact that there are 128
functional units on the SPARC T3-4, the chip multithreading is able to
improve the performance of our scenario-based DSE framework.

\section{Conclusions}\label{sec:conc}
\begin{figure}[!t]
\centering
\includegraphics[width=3in]{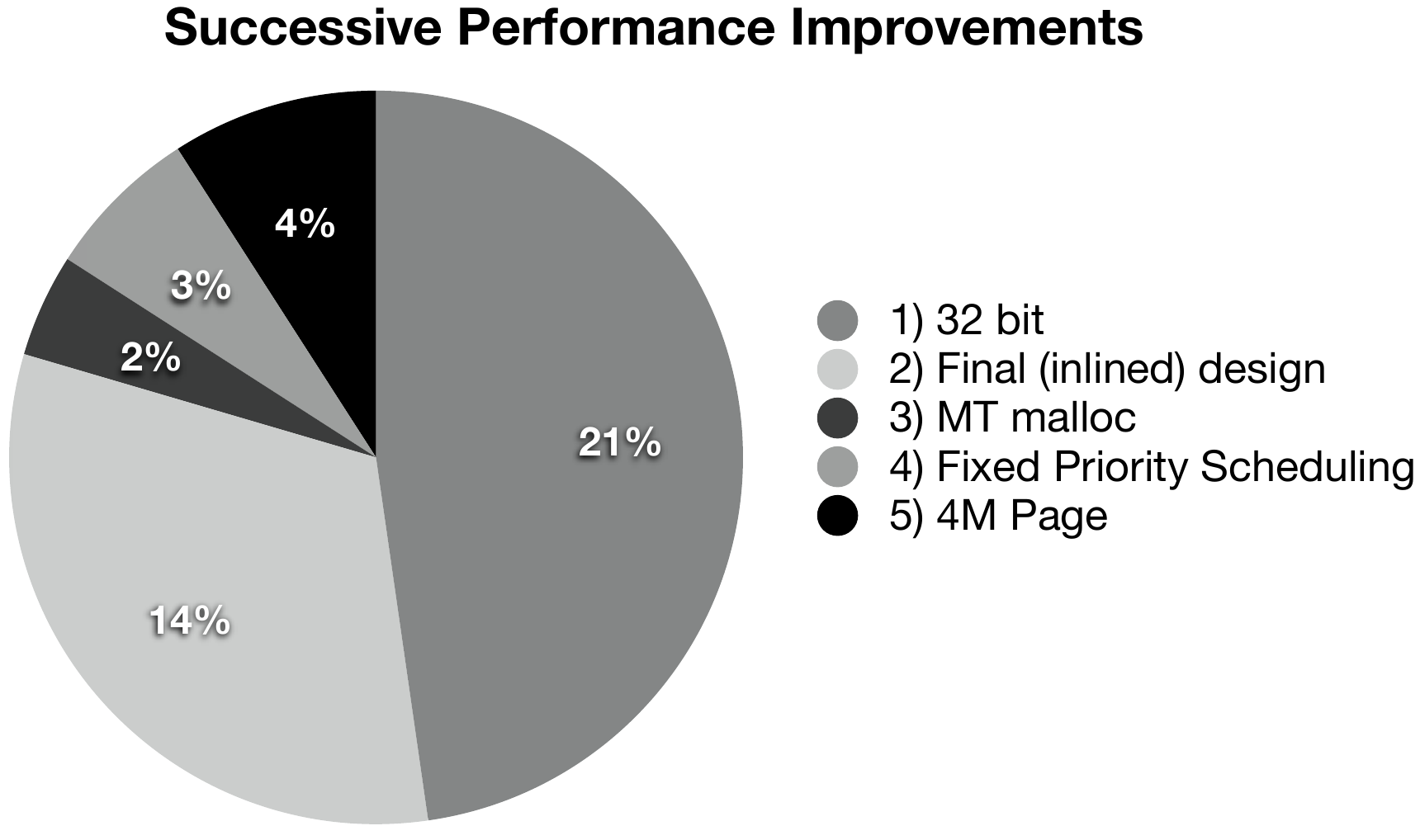}
\caption{The successive improvements in performance after the
different optimization steps. For each improvement step, the relative
execution time is given with respect to the plain 64 bit compilation.
Each of the improvement extends the improvements of the previous step.
As a consequence, at the 4M Page all the improvements are enabled (32
bit, final inlined design, MT Malloc and fixed priority scheduling)}
\label{fig:t3:improve}
\end{figure}

In this paper the porting of our scenario-based \ac{DSE} framework to
the SPARC T3-4 is described. In order to analyze the performance of
our scenario-based DSE on the SPARC T3-4, we used the Oracle Solaris Studio
Performance Analyzer to profile the application. This resulted in a
modified design of the scenario based \ac{DSE} where the locks in the shared job queue were replaced
by atomic operations. The only locks left in the application are
barrier synchronizations that are needed to ensure that the worker
threads do not access the queue when it is filled.

A summary of the improvements during the profiling of our
scenario-based \ac{DSE} framework on the SPARC T3-4 server is given in
Figure \ref{fig:t3:improve}. The largest improvement were made during
the first two steps. By using $32$ bits instead of $64$ bit
compilation, already $21$ percent performance improvement was
achieved. Secondly, the lockless queue implementation combined with
function inlining brought another $14$ percent of improvement. 
The heap allocation scheme and the scheduling policy, on the other hand, give moderate improvements (2 or 3\%)
on the final design of the scenario-based DSE framework. Largest remaining
bottleneck are the \ac{TLB} misses. A 4M pagesize already give a
performance improvement of more than 4\%, but additional gains could be achieved
in future work.

Finally, the SPARC T3-4 server gives a speedup of more than 185 times as fast
as the sequential code. Given the fact that there are only 128
execution units, we can conclude that the chip multithreading approach
is already paying off. Still, the SPARC T3-4 behaves poorly for a
workload with a large number of (similar) processes. When all the
hardware threads are filled, only 8 instruction \ac{TLB} entries are
available and 16 data \ac{TLB} entries. This can quickly give a
performance degradation by introducing \ac{TLB} misses.

\bibliography{literature}

\begin{thebibliography}{10}

\bibitem{oracleT3}
Sparc t3-4 server.
\newblock
  \url{http://www.oracle.com/us/products/servers-storage/servers/sparc-enterprise/t-series/sparc-t3-4-ds-173100.pdf}.

\bibitem{oracle4440}
Sun fire x4440 server.
\newblock
  \url{http://www.oracle.com/us/products/servers-storage/servers/x86/034679.pdf}.

\bibitem{gheorghita09}
S.~V. Gheorghita et~al.
\newblock System-scenario-based design of dynamic embedded systems.
\newblock {\em ACM Transactions on Design Automation of Electronic Systems},
  14(1):1--45, 2009.

\bibitem{graham04}
S.~L. Graham, P.~B. Kessler, and M.~K. McKusick.
\newblock gprof: a call graph execution profiler.
\newblock {\em SIGPLAN Notices}, 39(4):49--57, April 2004.

\bibitem{gries04}
M.~Gries.
\newblock Methods for evaluating and covering the design space during early
  design development.
\newblock {\em Integration, the VLSI Journal}, 38(2):131--183, 2004.

\bibitem{grove10}
D.~Grove.
\newblock {\em Multicore Application Programming: for Windows, Linux, and
  Oracle Solaris (Developer's Library)}.
\newblock Addison-Wesley Professional, 1st edition, November 2010.

\bibitem{jia10}
Z.J. Jia, A.D. Pimentel, M.~Thompson, T.~Bautista, and A.~Nunez.
\newblock Nasa: A generic infrastructure for system-level mp-soc design space
  exploration.
\newblock In {\em 8th IEEE Workshop on Embedded Systems for Real-Time
  Multimedia (ESTIMedia)}, pages 41--50, October 2010.

\bibitem{kahn74}
G.~Kahn.
\newblock The semantics of simple language for parallel programming.
\newblock In {\em IFIP Congress}, pages 471--475, 1974.

\bibitem{mitchell98}
M.~Mitchell.
\newblock {\em An Introduction to Genetic Algorithms}.
\newblock MIT Press, Cambridge, MA, USA, 1998.

\bibitem{pimentel06}
A.~D. Pimentel, C.~Erbas, and S.~Polstra.
\newblock A systematic approach to exploring embedded system architectures at
  multiple abstraction levels.
\newblock {\em IEEE Transactions on Computers}, 55(2):99--112, 2006.

\bibitem{stralen10c}
P.~van Stralen and A.~D. Pimentel.
\newblock {A High-level Microprocessor Power Modeling Technique Based on Event
  Signatures}.
\newblock {\em Journal of Signal Processing Systems}, 60(2):239--250, August
  2010.

\bibitem{stralen10b}
P.~van Stralen and A.~D. Pimentel.
\newblock Scenario-based design space exploration of {MPSoCs}.
\newblock In {\em Proceedings of IEEE International Conference on Computer
  Design (ICCD '10)}, October 2010.

\end{thebibliography}
\bibliographystyle{plain}

\end{document}